\documentclass[twocolumn,10pt,aps,prd]{revtex4-2}
\usepackage{hyperref,graphicx,amsfonts,amsmath,amssymb}
\usepackage[latin1]{inputenc}
\usepackage{multirow}
\usepackage{subcaption}
\usepackage{diagbox}
\newcommand{\tn}{\textnormal}
\newcommand{\GeV}{\tn{GeV}}
\newcommand{\TeV}{\tn{TeV}}
\begin{document}
\title{Charged Higgs decay to $W^{\pm}H$ at a high energy lepton collider}

\author{Majid Hashemi}
\email{hashemi_mj@shirazu.ac.ir}
\author{Laleh Roushandel}
\email{l.n.roushandel@gmail.com}
\affiliation{Physics Department, College of Sciences, Shiraz University, Shiraz, 71946-84795, Iran}

\begin{abstract}
In this work, we present a search strategy for heavy charged Higgs boson at Compact Linear Collider (CLIC) as a future $e^+e^-$ collider. The signal is charged Higgs boson pair production in two Higgs doublet model (2HDM) followed by $H^{\pm}\to W^{\pm}H$ and $H\to b\bar{b}$. Here, $H$ denotes the heavy CP-even neutral Higgs boson of the model. The collider center of mass energy is chosen to be $\sqrt{s}=1400$ GeV as the second stage of CLIC operation. In this case, $m_{H^+}<\sqrt{s}/2$ can be explored due to the pair production. It is shown that the signal of charged Higgs in the mass range $250~ \textnormal{GeV}<m_{H^+}< 650~ \textnormal{GeV}$ in fully hadronic final state, containing four $b$-jets from neutral Higgs and four jets from $W$ bosons, can well be observed on top of the standard model background. Finally $5\sigma$ contours are presented in $(m_{H^{\pm}},m_{H})$ space for different $\tan\beta$ values.        
\end{abstract}

\maketitle

\section{Introduction}
During the last decades, standard model of particle physics (SM) has been tested with a reasonable precision and it has proved to be a satisfactory model of electroweak and strong interactions. 

The Physics Nobel prize in 2013 was awarded to theoretical prediction of the Higgs boson \cite{Higgs1,Higgs2,Englert1,kibble1,Higgs3,kibble2} which had been experimentally confirmed by the two CMS and ATLAS collaborations of the Large Hadron Collider (LHC) experiment \cite{HiggsCMS,HiggsATLAS}.

Soon after observation of the new boson, extensive studies focused on its properties including production cross section, decay rates, couplings, spin, parity and CP structure \cite{LHC1,LHC2,LHC3,LHC4,LHC5,LHC6,LHC7,LHC8}. There is overall agreement between the measured properties of the new boson with SM prediction. 

There are, however, open problems which imply that the underlying theory of nature is beyond SM. There is a long list of such problems. One of the main issues is the so called hierarchy problem, i.e., finite mass of the observed Higgs boson which is theoretically sensitive to radiative corrections and the large difference between the electroweak scale and the Planck mass. One natural solution for the hierarchy problem is supersymmetry which requires an extended Higgs sector (at least two Higgs doublets) and reduces the divergence of radiative corrections to the Higgs boson mass without a fine tuning of the model parameters \cite{HTPH, habersusy}. 

Another main issue is the origin of dark matter (DM) discussed in a variety of theoretical models such as inert doublet model (IDM) \cite{IDM0,IDM1,IDM2,IDM3,IDM4,IDM5,IDMCS}, SM with a real \cite{RealxSM1,RealxSM2,RealxSM3,RealxSM4,RealxSM5,RealxSM6,RealxSM7} or complex scalar singlet \cite{Complex1,Complex2,Complex3,Complex4,Complex5} and supersymmetry \cite{SUSYDM}. The collider searches for the dark matter candidates are one of the main tasks in parallel with searches for extra Higgs bosons and supersymmetry \cite{ColliderDM1,ColliderDM2}.       

Other scenarios address the CP violation through baryon asymmetry at the electroweak scale \cite{CPViolation} or the neutrino mass \cite{NuMass}.
 
These beyond SM (BSM) scenarios all need extended Higgs sectors and can be built to predict and confirm the observed particle at LHC. In such a situation, the observed particle belongs to a family of Higgs bosons predicted in a more general model.

One of the most attractive BSM candidates is two Higgs doublet model (2HDM) which has been considered as a model of CP violation in its original form \cite{2hdm1}. There is currently a high motivation for this model not necessarily as a basis for supersymmetry.

The Higgs sector of 2HDM incorporates two Higgs doublets resulting in five physical Higgs bosons \cite{2hdm1,2hdm2,2hdm3}, i.e., three neutral bosons ($h,~H,~A$) and two charged bosons ($H^{\pm}$) \cite{2hdm_HiggsSector1,2hdm_HiggsSector2}. 

The recently observed deviation of the $W$ boson mass from SM prediction by the CDF collaboration \cite{Wmass} has imposed constraints on the 2HDM parameter space. The deviations in $m_W$ had been explained in terms of singlet extension to SM \cite{Wmass1} and 2HDM \cite{Wmass2} before CDF announcement. However, there has been a large number of recent studies discussing the charged and neutral scalar mass splitting \cite{DeltamH1,DeltamH2,DeltamH3,DeltamH4,DeltamH5, WmassTwoLoop}, and the interplay between the 2HDM and the muon $g-2$ anomaly in the light of the recent $m_W$ measurement \cite{mWg20,mWg21,mWg22,mWg23,mWg24}. There are also studies which find upper limits of 1 TeV on the masses of the charged and neutral 2HDM scalars \cite{mWupperlimit,mWupperlimit2}. 

In a CP conserving scenario without FCNC, there are four types of 2HDM in terms of Higgs-fermion diagonal couplings with a rich phenomenology \cite{Barger_2hdmTypes,2hdm_TheoryPheno}. The type 2 serves as the basis to build the Higgs sector of the Minimal Supersymmetric SM (MSSM) \cite{MSSM1,MSSM2,MSSM3,MSSM4}. 

In the so called alignment limit, one of the 2HDM neutral Higgs bosons (usually the lightest) acquires the same properties as those of the SM Higgs at tree level and plays the role of the LHC observation at 125 GeV \cite{align1,align2,align3,m12}. The tree level alignment is however broken when loop corrections are included. Other 2HDM neutral Higgs bosons can have different properties (decay rates, CP, ...) which makes them distinguishable from the lightest boson especially at the decoupling limit which occurs if the mass difference between the SM-like Higgs boson ($h$) and heavy bosons is large \cite{decoupling,pos5}.

Observation of a new scalar is a signal for BSM with extended Higgs sector. The signal of these scalars can be different in terms of their decay products and decay rates. If one of these scalars has SM-like couplings, the other one has couplings which are significantly different from the SM Higgs (e.g., no coupling to W and Z bosons). On the other hand, there is no elementary charged scalar in SM and observation of a charged Higgs with its unique signatures (electric charge, decay channels, ...) is a crucial proof that the underlying theory is beyond SM.

The charged Higgs bosons were used to explain the observed deviations of flavor physics observables from SM predictions in the past. These observables were $b\to s\gamma$ transition rate \cite{Misiak,Misiak2,Misiak3,Misiak4}, branching ratio of heavy meson decay to $\tau\nu$ including $B_u \to \tau\nu$, $D_s \to \tau\nu$ \cite{BMeson1} and $B_s \to \mu^+\mu^-$ \cite{BmmLHCb,BmmCMS,BmmATLAS}. These decays all contain diagrams involving charged weak currents which yield different results if charged Higgs boson contributions are added. A fine tuning of the model parameters achieves an agreement between 2HDM prediction and experimental observations better than SM \cite{FMahmoudi}. The observed 2.4$\sigma$ deviation in fully leptonic decays of $B$ mesons reported in 2020 has recently reduced to less than 1$\sigma$ below SM \cite{BmmLHCb:2021,BmmCMS:2022}. Even in case of no deviation, these measurements impose strong constraints on 2HDM parameters \cite{FMahmoudi2,FMahmoudi3}. Currently $b\to s\gamma$ channel provides the strongest lower limit on $m_{H^+}$ \cite{Misiak,Misiak2,Misiak3,Misiak4}. These studies are considered as indirect searches for new physics and are complementary to the direct searches at colliders \cite{FMahmoudi4}. 

Here we are going to focus on the Higgs sector of 2HDM and provide collider signatures which can be explored at future colliders for charged Higgs boson discovery. In what follows, a review of theoretical framework and collider searches for the charged Higgs and their current results is presented. After a discussion on different decay channels of the neutral and charged Higgs bosons, the charged Higgs decay to $W^{\pm}$ boson and a heavy neutral Higgs boson is introduced as the search channel following previous studies using $H^+ \to t\bar{b}$ and $H^+ \to \tau^+\nu$. We present the analysis as a proposal for a future high energy $e^+e^-$ collider which is specifically assumed to be CLIC in its second stage of operation \cite{cliccdr}. We show that heavy charged Higgs mass region can well be probed at lepton colliders through $e^+e^- \to H^+H^-$ in the fully hadronic final state. Results show reasonable performance of such colliders compared to LHC.  
\section{Theoretical Framework}	
\label{thfr}
The Higgs sector of the 2HDM Lagrangian is a natural expansion of SM to contain two complex scalar doublets with kinetic terms and the potential written in the form:
\begin{eqnarray}\label{lagrangian-2HDM}
	\mathcal L_{\Phi}^{\textnormal{2HDM}}&=& \sum_{i=1,2}(D_{\mu}\Phi_{i})^{\dagger}(D^{\mu}\Phi_{i})-\mathcal V^{\textnormal{2HDM}}.
\end{eqnarray}
The Higgs-gauge interaction terms are embedded in the kinetic term containing $D_\mu$ (the covariant derivative) and the two doublets ($\Phi$) are 
\begin{eqnarray}\label{phi2HDM}
	\Phi_{i}=\binom{\phi_{i}^{+}}{(v_{i}+\rho_i+i\eta_i)/\sqrt{2}},~ i=1, 2.  
\end{eqnarray}
The neutral and charged Higgs fields are obtained by introducing two mixing angles $\alpha$ and $\beta$ acting on the neutral and charged parts of the doublet containing $\rho_i$, $\eta_i$ and $\phi_i^{\pm}$ through  
\begin{align}
	h&=-\rho_1\sin\alpha+\rho_2\cos\alpha \nonumber\\ 
	H&=\rho_1\cos\alpha+\rho_2\sin\alpha \nonumber \\ 
	A&=\eta_1\sin\beta+\eta_2\cos\beta  \nonumber\\ 
	H^{\pm}&=\phi^{\pm}_1\sin\beta+\phi^{\pm}_2\cos\beta.
	\label{mixing}
\end{align}
The ratio of vacuum expectation values of the two doublets is related to the mixing angle $\beta$ through $\tan\beta=v_2/v_1$ under the conditions $v_1^2+v_2^2=v^2=(246 ~\textnormal{GeV})^2$. At the alignment limit which occurs if $\sin(\beta-\alpha)=1$ or $\beta-\alpha=\pi/2$, the $\beta$ parameter determines the Higgs-fermion couplings \cite{tanbsignificance}.
 
The Higgs boson mass terms and Higgs self-interactions are formulated in the potential term which is written as follows:
\begin{align}
	\mathcal{V}^{\textnormal{2HDM}} \nonumber &= m_{11}^2\Phi_1^\dagger\Phi_1+m_{22}^2\Phi_2^\dagger\Phi_2-m_{12}^2\left(\Phi_1^\dagger\Phi_2+\Phi_2^\dagger\Phi_1\right)\\
	\nonumber &+\frac{1}{2}\lambda_1\left(\Phi_1^\dagger\Phi_1\right)^2+\frac{1}{2}\lambda_2\left(\Phi_2^\dagger\Phi_2\right)^2\\ \nonumber
	\nonumber &+\lambda_3\left(\Phi_1^\dagger\Phi_1\right)\left(\Phi_2^\dagger\Phi_2\right)+\lambda_4\left(\Phi_1^\dagger\Phi_2\right)\left(\Phi_2^\dagger\Phi_1\right)\\ \nonumber
	\nonumber &+\frac{1}{2}\lambda_5\left[\left(\Phi_1^\dagger\Phi_2\right)^2+\left(\Phi_2^\dagger\Phi_1\right)^2\right].\\ 
\label{potential}
\end{align}
This form of the potential (up tp a soft symmetry breaking term $m_{12}$) respects the $Z_2$ symmetry, i.e., invariance of the Lagrangian under interchange of $\phi_1 \to \phi_1,~\phi_2 \to -\phi_2$ or $\phi_1 \to -\phi_1,~\phi_2 \to \phi_2$ which prevents $\phi_1 \rightleftarrows \phi_2$ transitions which result in CP violation \cite{uni3}. Therefore the above form of the potential is the basis for a CP conserving 2HDM. 

The Higgs-fermion couplings can be separated to the neutral Higgs interactions with fermions and the charged Higgs sector. The former has been studied in recent analyses extensively including our works on type 1 \cite{HG_2,HG_3,HG_4,HE}, type 3 \cite{HG_1,HN,HM} and type 4 \cite{MH_1,MH_2,HG_5,HM}. In this analysis we focus on the charged Higgs search at a lepton collider with enough center of mass energy to explore high masses which have been out of LHC reach. 

The charged Higgs interaction with fermions takes the following Lagrangian:    
\begin{align}
	\mathcal{L} = \frac{g}{2\sqrt{2}m_W} H^{\pm}& [ V_{ij} m_{u_i} \rho^u \bar{u}_i(1 - \gamma^5) d_j \nonumber\\
	&+ V_{ij} m_{d_j} \rho^d \bar{u}_i(1 + \gamma^5) d_j \nonumber\\
	&+ m_l \rho^l \bar{\nu}_i (1 +\gamma^5)l_i ] + h.c.
	\label{lag}
\end{align}
where $v=2m_W/g$, $V_{ij}$ are CKM matrix elements and coupling factors $\rho$ are given in Tab. \ref{hf} for each type of the 2HDM.  
\begin{table}[h]
	\centering
	\begin{tabular}{|c|ccc|}
		\hline
		\multirow{2}{*}{
			\diagbox{Model type}{Coupling}} & $\rho^u$& $\rho^d$& $\rho^l$\\
		& & & \\
		\hline
		1 & $\cot\beta$ & $\cot\beta$ & $\cot\beta$  \\
		\hline
		2 & $\cot\beta$ & $-\tan\beta$ & $-\tan\beta$ \\
		\hline
		3 & $\cot\beta$ & $-\tan\beta$ & $\cot\beta$ \\
		\hline
		4 & $\cot\beta$ & $\cot\beta$ & $-\tan\beta$ \\
        \hline
	\end{tabular}
	\caption{Higgs-fermion couplings in different types of 2HDM at the alignment limit.}
	\label{hf}
\end{table}

There are also charged Higgs decay to gauge bosons in the form of $H^{\pm} \to W^{\pm}\phi$ with $\phi=h/H/A$ as one of the neutral Higgs bosons. The vertices and decay rates of these channels are type independent as listed below:
\begin{align}
	HH^+W^-&:\frac{g\sin(\beta-\alpha)}{2}\nonumber\\ 
	hH^+W^-&:\frac{g\cos(\beta-\alpha)}{2}\nonumber\\ 
	AH^+W^-&:\frac{ig}{2}. 
	\label{HW}
\end{align}

However, the branching ratios depend on the 2HDM type due to dependence of other fermionic decay channels. At the alignment limit, with $\sin(\beta-\alpha)=1$, $H^+ \to W^+h$ is suppressed while other decays. i.e., $H^+ \to W^+H$ and $H^+ \to W^+A$ can be significant if two conditions are satisfied. 

The first condition is that the fermionic decays are suppressed. This suppression occurs for heavy charged Higgs decay $H^+ \to t\bar{b}$ at moderate $\tan\beta$ values which minimize the term $m_t\cot\beta+m_b\tan\beta$ for the $H^+\bar{t}b$ coupling in Eq. \ref{lag}. This condition occurs in 2HDM types 2 and 3 at $\tan\beta=\sqrt{m_t/m_b}\simeq 6.5$.  Other types disfavor $H^+ \to t\bar{b}$ at high $\tan\beta$ values. 

The second condition is to have enough kinematic phase space for the decay. In MSSM-like 2HDM scenarios with degenerate masses for the Higgs bosons, these Higgs conversion decays are kinematically forbidden. It has been shown that for $\Delta\rho$ to be small enough and consistent with electroweak precision measurements, at least one of the heavy neutral Higgs bosons should have the same mass as the charged Higgs boson, i.e., $m_H=m_{H^+}$ or $m_A=m_{H^+}$ \cite{drho1,drho2,drho3,drho4}. The second option is adopted in this analysis leading to the mass spectrum $m_h<m_H<m_A(=m_{H^+})$. Therefore $H^+ \to W^+A$ is naturally suppressed and the only remaining non-fermionic decay is $H^+ \to W^+H$. We will describe the analysis of this channel in details after a brief review of the past and recent searches for the charged Higgs boson in different decay channels in the next section.

\section{Review of the current results}  
The charged Higgs searches have been historically divided into two main categories of light ($m_{H^+} \lesssim m_t$) and heavy ($m_{H^+} \gtrsim m_t$) regions. Each region has its own characteristics which is described briefly.
\subsection{Light charged Higgs}
The light charged Higgs can be produced in top quark decay through $t \to H^+b$ and competes with SM decay $t\to W^+b$. In the absence of charged Higgs boson, the top quark branching ratio of decay to $W$ is close to unity. However, in certain circumstances, e.g., when the coupling factor, $\rho^d$ in the second term of Lagrangian (Eq. \ref{lag}) is large, $t \to H^+b$ decay rate can be significant. This situation occurs in 2HDM types 2 and 3 where $\rho^d=-\tan\beta$ and enhances at high $\tan\beta$ values. Since the main decay channels in this case are $t\to W^+ b$ and $t \to H^+ b$ and sum of their branching ratios has to be unity, enhancement of $(t\to H^+ b)$ decay rate as $\tan^2\beta$ leads to suppression of $(t \to W^+b)$. One has to note that enough phase space is also required for significant decay rate and $m_{H^+}$ values close to the top quark mass result in suppression of non-SM decay $t \to H^+b$. When the light charged Higgs is produced in top quark decay, usually $H^+ \to \tau^+\nu$ is analyzed because $H^+ \to t\bar{b}$ is kinematically suppressed. The $\tau$ lepton in its hadronic decay produces a sharp signature for the charged Higgs boson due to the mass difference between $H^+$ and $W^+$, and spinless nature of the charged Higgs boson compared to the case of spin-1 $W$ boson. These effects have been discussed extensively in the literature \cite{CHOld,CHDPRoyFermilab,taupol,CHRoy1,CHRoy2,CHRoy3,CHSharpening}. 

The charged Higgs boson production through the top quark decay ($t \to H^+b$) and its decay to $\tau$ jet ($H^+ \to \tau^+\nu$) has been the search channel from the time of LEP \cite{lep1,lep2,lepexclusion1} to TeVatron collaborations D0 \cite{d01,d02,d03,d04} and CDF \cite{cdf1,cdf2,cdf3}. The preliminary studies at LHC \cite{MyLightCH,BiscaratLightCH} were followed by CMS \cite{LightCH7TeVCMS1,LightCH7TeVCMS2} and ATLAS \cite{LightCH7TeVATLAS1,LightCH7TeVATLAS2} using 7 TeV data. The 8 TeV results were then reported by CMS \cite{CHtaunu8TeVCMS} and ATLAS \cite{CHtaunu8TeVATLAS}. Currently the light charged Higgs masses below 160 GeV are excluded for all $\tan\beta$ values in MSSM (which is based on 2HDM type 2) as reported by 13 TeV LHC data analyses \cite{CMS:chnew6,ATLAS:chnew1}.          
\subsection{Heavy charged Higgs}  

If the charged Higgs boson is heavier than the top quark, it can not be produced in on-shell top quark decays. The off-shell production of the top quarks can instead be used to produce the charged Higgs bosons. The light charged Higgs production process which is $pp \to t\bar{t}$ at LHC, with at least one of the top quarks decaying to the charged Higgs, is replaced by $pp \to t\bar{b}H^-$ whose cross section calculation was performed in several works before the LHC startup \cite{gbCH1,gbCH2,gbCH3,gbCH4,gbCH5}. 

In the heavy charged Higgs region there is still possibility to adopt $H^+ \to \tau^+\nu$ for the search, especially in 2HDM types 2 and 4 which allow enhancement of the charged Higgs leptonic decay with its decay rate being proportional to $\tan^2\beta$. 
             
The alternative decay channel is $H^+ \to t\bar{b}$ which benefits from a larger mass term in the vertex ($m_b$ vs. $m_\tau$) and can be significant in 2HDM types 2 and 3 with the same reasons as discussed about $t \to H^+b$ previously. This channel has already been studied in the literature \cite{KAHeavyCH1,KAHeavyCH2,st2,jss8,tbSuperCollider,HeavyCHatLHC,CH3b,Hashemi:2015hxc} followed by LHC results reported by ATLAS \cite{ATLAS:chnew5,ATLAS:chnew2} and CMS collaborations \cite{CMS:chnew1,CMS:chnew4} which excluded charged Higgs masses above 200 GeV for $\tan\beta$ values below 2.1 and also above 34.

The analysis of the charged Higgs decay to $c\bar{s}$ has also been recently reported by CMS collaboration \cite{CMS:chnew3}. The off-diagonal decay to $c\bar{b}$ has been analyzed by CMS \cite{CMS:chnew7} and ATLAS collaborations \cite{ATLAS:chnew3,ATLAS:chnew4} with their results reported as upper limits on the product of production cross section times branching ratio of the light charged Higgs decay.  

Other decay channels have also been proposed such as $H^+ \to \mu^+\nu$ \cite{Hashemi:2011gy,CH2munuLHC}.    

It should be noted that the final state of $H^+ \to t\bar{b}$ with $t\to W^+b$ and $W\to jj$ is the same as $H^+ \to W^+H$ followed by $W\to jj$ and $H\to b\bar{b}$. These two modes can be comparable with large branching ratios in the region near $\tan\beta \sim 7$ as discussed extensively in \cite{CH2WhWH}. 

In the context of 2HDM type 1, the charged Higgs bosonic decays have been studied in \cite{CH2WhArhrib}. The charged Higgs-$W$ boson coupling involved in $pp\to H^{\pm}\phi$ production at LHC with $\phi=h/H/A/W^{\mp}$ together with decays with the same vertices involved are discussed in \cite{CH2WH} with several benchmark scenarios introduced for LHC searches. The same type of charged Higgs associated production with $\phi=h/A$ or $H^+H^-$ pair production with $H^{\pm}\to W^{\pm}h/A$ have been discussed in \cite{CHLightNewModes} and the regions of parameter space where the charged Higgs bosonic decays are enhanced are identified. 

An analysis of $H^{\pm}\to W^{\pm}h_{125}$ has been proposed for LHC in \cite{CH2WhLHC}. The loop contributions to $H^{\pm} \to W^{\pm}V$ ($V=\gamma/Z$) has been calculated in \cite{CH2WLoop} leading to one to three orders of magnitude enhancement in those branching ratios. A proposal for the energy upgrade of LHC to probe $H^{\pm}W^{\mp}Z$ interaction has been presented in \cite{CHWZ}. 

The CMS collaboration has reported their first results of heavy charged Higgs search by analyzing $H^+ \to W^+H$ \cite{CMS:chnew2}. They have set upper limits on the cross section times BR($H^+ \to W^+H$) for the charged Higgs masses above 300 GeV and the heavy CP-even neutral Higgs mass fixed at $m_H=200$ GeV. The analysis of $H^+ \to W^+A$ has also been reported for a light charged Higgs in the mass range $100 ~\GeV<m_{H^+}<160 ~\GeV$ \cite{CMScH2WA,CMScH2WA2}. 

The $H^+ \to W^+Z$ has been searched for by CMS \cite{CMScH2WZ} and ATLAS \cite{ATLAScH2WZ,ATLAScH2WZ2} collaborations. The multi-lepton final state of the charged Higgs decay to vector boson has been reported in \cite{ATLASCH2W}.

Despite the extensive search for the charged Higgs bosonic decays, a large region of the parameter space is still unexplored and will be shown to be out of HL-LHC reach too.

\section{Analysis of $H^+ \to W^+H \to jjbb$ at CLIC}
\subsection{Collider choice}
There are different scenarios for the future of colliders in hadron-hadron and lepton-lepton collision modes. The Large Hadron Collider (LHC) after an upgrade is going to operate in high luminosity mode HL-LHC \cite{hllhc1,hllhc2}. The new design as Future Circular Collider is going to be the next generation collider operating in two modes of hadronic collisions with $\sqrt{s}=100~\TeV$ as FCC-hh \cite{FCC-hh}, and $e^+e^-$ collisions with $\sqrt{s}=350~\GeV$ as FCC-ee in top factory mode \cite{FCC-ee1,FCC-ee2}. There are other proposals such as International Linear Collider (ILC) \cite{ILC,ILCEnergy,ILCEnergy2} operating at $\sqrt{s}=500~\GeV$ and CEPC \cite{CEPC1,CEPC2} with $\sqrt{s}=240 ~\GeV$. 

The above hadron collision programs will follow LHC in their luminosity and energy frontier \cite{FCCEnergy}. In lepton collision modes there have been numerous studies of lepton colliders and their potential for Higgs boson searches  at ILC \cite{ilchiggs} and CEPC \cite{cepchiggs}. However, a suitable choice for heavy Higgs boson searches is Compact Linear Collider (CLIC) in its high energy operation modes (stages 2 and 3 with $\sqrt{s}=1400$ and 3000 GeV respectively) \cite{clichiggs1,clichiggs2}. The first stage of CLIC operation will be at $\sqrt{s}=350~\GeV$ and can not be used for heavy Higgs boson searches targeted in this work. Therefore the first realization of a lepton collider providing the opportunity to search for charged Higgs bosons in pair production with masses above 200 GeV occurs at the second stage of CLIC at $\sqrt{s}=1400~\GeV$. This is the collider choice for the analysis which will be described in this paper. The ultimate operation mode can be reserved to focus on unexplored regions of the 2HDM parameter space which escaped from the previous stages.         
\subsection{Software setup}
The branching ratios of neutral and charged Higgs bosons as well as theoretical constraints are obtained using \texttt{2HDMC-1.8.0} \cite{2hdmc1,2hdmc2,2hdmc3}. The experimental exclusion regions are obtained using \texttt{HiggsTools-1} \cite{higgstools}. The cross section and event generation is performed with the use of \texttt{WHIZARD-3.1.2} \cite{whizard1,whizard2} which benefits from beam spectra for lepton colliders using sub-package \texttt{circe2} \cite{circe2}. Events containing hard scattering are passed to \texttt{PYTHIA-8.3.09} \cite{pythia} for multi-particle interactions and final state radiation and showering. The detector simulation is performed using \texttt{DELPHES-3.5.0} \cite{delphes1,delphes2,delphes3}. The detector card \texttt{CLICdet$\_$Stage2} is used to perform the physical object reconstruction including jet reconstruction and flavor association, $b$-tagging (in 90$\%$ efficiency scenario) including fake rate ($p_T$ and $\theta$ dependent), charged track and jet momentum smearing \cite{clicdp,overlay,clicdet}. The final result visualization is done using \texttt{ROOT-6.28} \cite{root} and \texttt{python3} libraries \texttt{numpy} \cite{numpy} and \texttt{matplotlib} \cite{matplotlib}.
\subsection{Theoretical constraints}
The Higgs sector potential in Eq. \ref{potential} is subject to theoretical requirements of positivity (being bounded from below) \cite{pos1,pos2,pos3,pos4,pos5}, unitarity (of the scattering S-matrix) and perturbativity (in the Higgs quartic interactions) \cite{uni1,uni2,uni3} and $\Delta\rho$ (to be within the range of electroweak precision measurements) \cite{drho1,drho2,drho3,drho4}. 

The benchmark points considered in this work fall in the range $250 ~\GeV < m_{H^+} < 650~ \GeV$. The signal cross section with heavier charged Higgs bosons tends to zero when reaching $m_{H^+}\to\sqrt{s}/2$. For each charged Higgs mass, the neutral CP-even Higgs boson mass is selected from the range $150 ~\GeV < m_H < 550 ~ \GeV$. Both mass increments are 100 GeV and the minimum mass difference between $m_H$ and $m_{H^+}$ has to be 100 GeV. The CP-odd Higgs boson mass is equal to the charged Higgs mass as stated in section \ref{thfr}. The analysis is presented for $\tan\beta=10$ which is close to the excluded region of neutral Higgs boson in 2HDM type 3.

The above theoretical requirements are satisfied for all points except for the point $(m_{H^+},m_H)=(650,150)~\GeV$ where only positivity, unitarity and $\Delta \rho$ requirements are satisfied. Although the signal process does not involve $2\to2$ Higgs interactions which are verified by perturbativity requirement, all theoretical considerations have to be taken into account for the potential to be physically sensible. However, we keep this point in the list for completeness of the search domain which is available at this center of mass.   
\subsection{Neutral Higgs boson considerations}
The current analysis is going to focus on the heavy charged Higgs search through $H^+ \to W^+H$. Therefore it is more relevant to (but still different from) the CMS analysis reported in \cite{CMS:chnew2} where the final state is obtained via $H^+ \to W^+H$ followed by $H\to \tau\tau$ with at least one of the $\tau$ leptons decaying hadronically. It is based on obtaining the distribution of charged Higgs transverse mass after a reasonable treatment of the missing transverse energy originated from hadronic decays of $\tau$ leptons, $\tau$ tagging as well as top quark tagging as the production process in their case is $pp \to t\bar{b}H^-$. 

\begin{figure}[hbt!]
	\centering
	\includegraphics[width=\linewidth,height=0.85\linewidth]{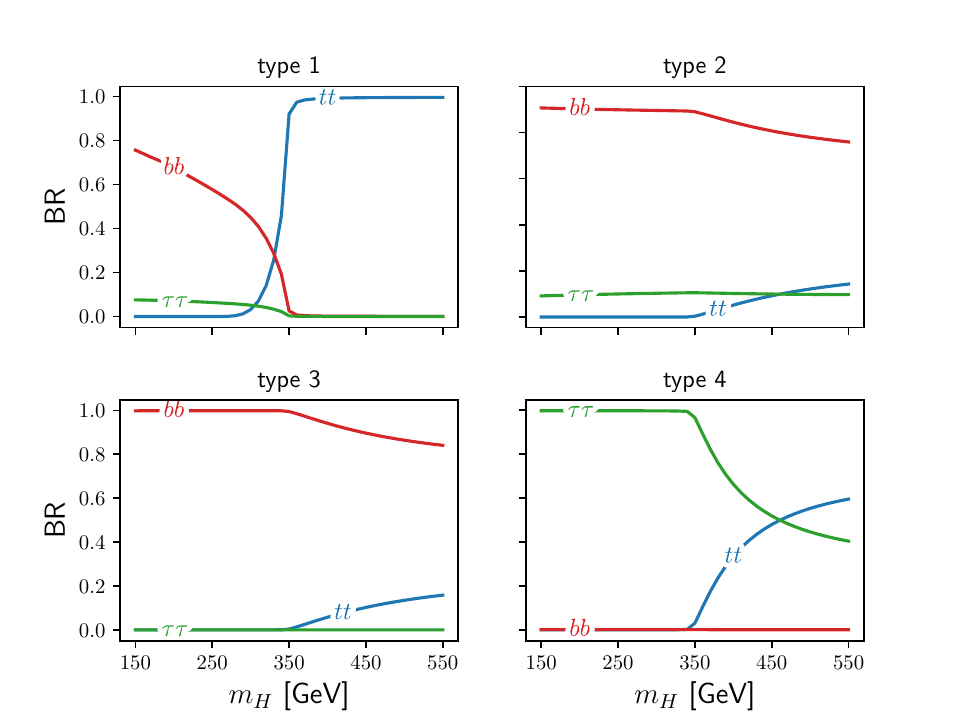}
	\caption{The neutral Higgs boson branching ratio of decay to different final states $b\bar{b}, \tau^+\tau^-$ and $t\bar{t}$ as a function of the Higgs boson mass at $\tan\beta=10$ in different types of 2HDM.}
\label{Hdecays}
\end{figure}

\begin{figure}[hbt!]
	\centering
	\includegraphics[width=\linewidth,height=0.85\linewidth]{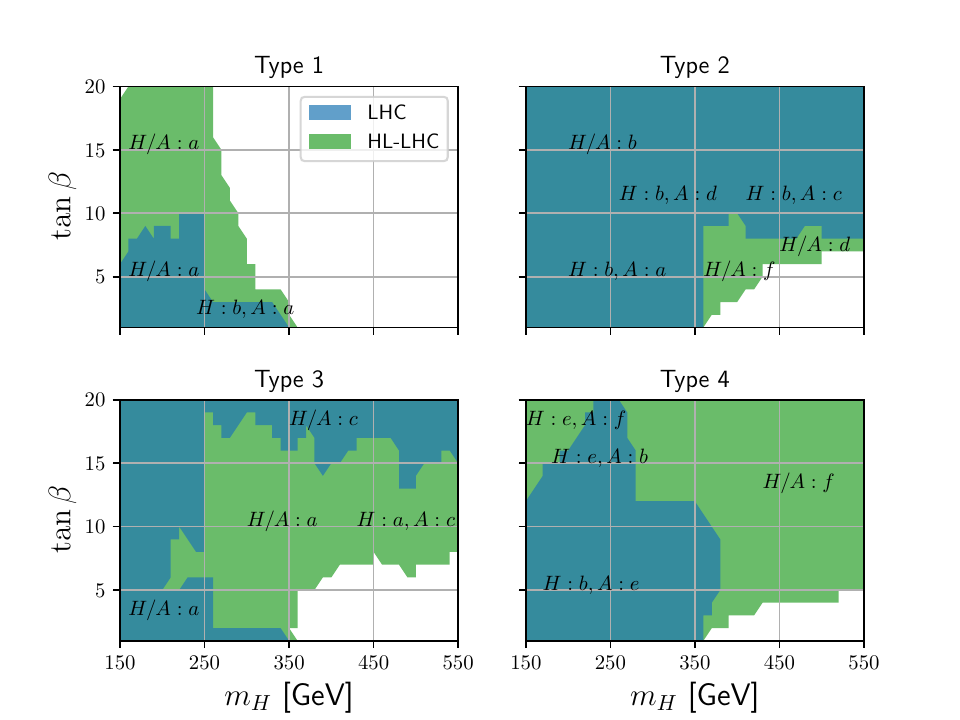}
	\caption{The 95$\%$ C.L. excluded regions of the heavy neutral Higgs bosons shown in blue (LHC) and green (HL-LHC expectation) as a function of the Higgs boson mass vs. $\tan\beta$. Labels show the corresponding analyses with highest sensitivity for CP-even H and CP-odd A exclusion: a: $A\to ZH$ \cite{atlas1}, b: $\phi \to \tau\tau$ \cite{T4_1}, c: $jj\gamma$ \cite{T3_1}, d: $\phi \to \tau\tau$ \cite{T2}, e: $H/A \to Z A/H$ \cite{T4_2}, f: $\phi \to \tau\tau$ \cite{MSSMHtautau} ($\phi=h/H/A$).}
\label{HExclusion}
\end{figure}

These event characteristics are different from what we are going to propose in the current work. In fact, in 2HDM types 1 and 3, the relevant decay mode for the heavy neutral Higgs boson is $H \to b\bar{b}$ as shown in Fig. \ref{Hdecays}. In type 1, $H\to b\bar{b}$ is suppressed if $m_H\gtrsim 2m_t$ but in that region the dominant decay channel is $H\to t\bar{t}$ which produces a large final state particle multiplicity and a signal discrimination from the background will be challenging. The other 2HDM types 2 and 4 are highly excluded by the current experimental searches at LHC or will be covered at high luminosity LHC (HL-LHC) as shown in Fig. \ref{HExclusion}. 

The HL-LHC exclusion region expectation is obtained by statistical extrapolation of current results from CMS and ATLAS to integrated luminosity of $\mathcal{L}_{\tn{HL-LHC}}=3000 fb^{-1}$ per experiment leading to total $\mathcal{L}=6000 fb^{-1}$ \cite{hllhc2}. In order to do so, the current signal ratios in \texttt{HiggsBounds} datasets are scaled by a factor of $\sqrt{\mathcal{L}_{\tn{current}}/\mathcal{L}_{\tn{HL-LHC}}}$ where the current integrated luminosity is taken from the dataset file. 

According to Fig. \ref{HExclusion}, lower limits of 250 and 350 GeV are set on the neutral Higgs masses based on the current LHC exclusion and HL-LHC expectation for both 2HDM types 1 and 3 with $\tan\beta<10$.

The analysis to be presented is limited to 2HDM types 1 and 3 in 4$b$ final state. The 2HDM type 4 is left due to the possibility of coverage by HL-LHC and lower efficiency of $\tau$-tagging compared to $b$-tagging. The $\tau$-tagging efficiency at CLIC in the transverse momentum range $50~ \GeV < p_T < 125~\GeV$ is $\sim60\%$ \cite{clictau}, while the $b$-tagging efficiency can be as high as 90$\%$ \cite{clicb}.        
\subsection{Charged Higgs boson considerations}
\begin{figure}[hbt!]
	\centering
	\includegraphics[width=\linewidth,height=0.85\linewidth]{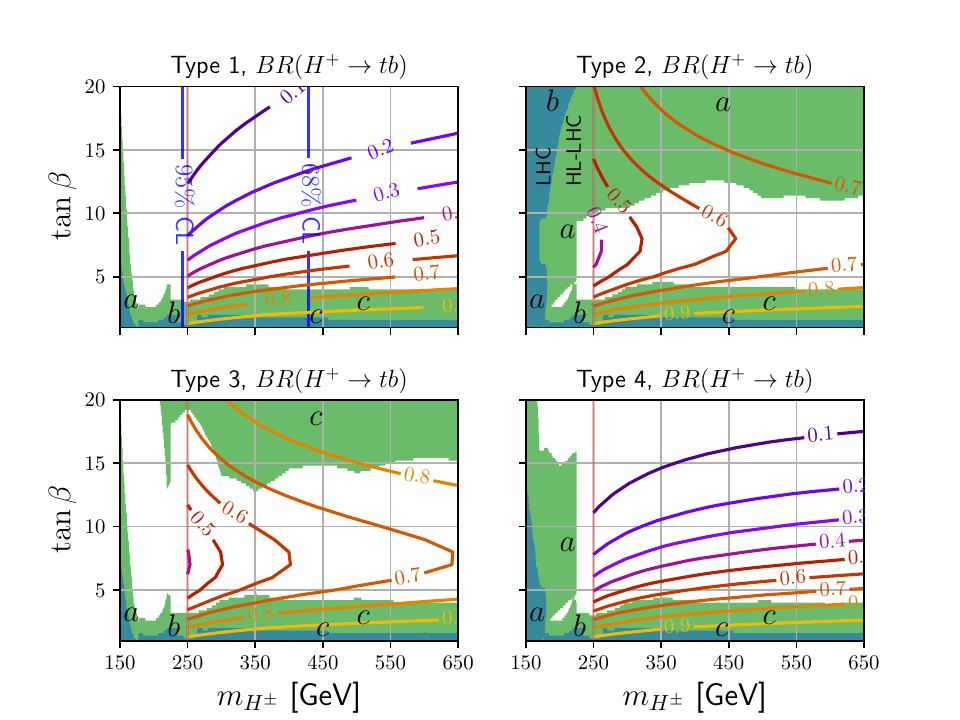}
	\caption{The charged Higgs branching ratio of decay to $t\bar{b}$ as a function of the mass and $\tan\beta$ in different types of 2HDM. The blue and green regions show the LHC 95$\%$ C.L. excluded region and HL-LHC expectation respectively. Labels show the corresponding analyses with highest sensitivity: a: ATLAS $H^+ \to \tau\nu$ \cite{ATLAS:chnew1}, b: ATLAS $H^+ \to t\bar{b}$ \cite{ATLAS:chnew2} and c: CMS $H^+ \to t\bar{b}$ \cite{CMS:chnew1}. The neutral Higgs boson mass is set to $m_H=m_{H^+}-100$ GeV. The vertical blue lines show the lower limit of the charged Higgs boson mass within 68$\%$ and 95$\%$ CL bound from $h_{125} \to \gamma\gamma$ measurement at LHC.}
\label{CH2tb}
\end{figure}

\begin{figure}[hbt!]
	\centering
		\includegraphics[width=\linewidth,height=0.85\linewidth]{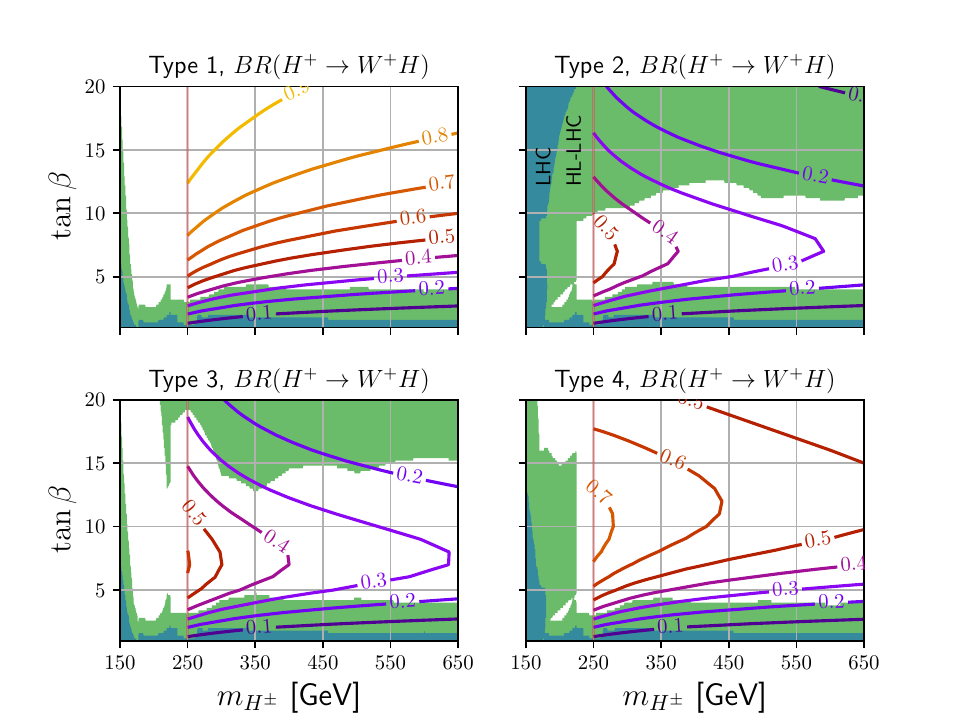}
	\caption{The charged Higgs branching ratio of decay to $WH$ as a function of the mass and $\tan\beta$ in different types of 2HDM. The neutral Higgs boson mass is set to $m_H=m_{H^+}-100$ GeV.}
\label{CH2WH}
\end{figure}

The charged Higgs boson is pair produced at $e^+e^-$ collisions through $e^+e^- \to \gamma/Z/h/H \to H^+H^-$. The contribution from neutral Higgs bosons in the $s$-channel propagator is negligible due to the low Higgs-electron coupling. The process is thus effectively Drell-Yan event (with incoming $e^+e^-$) producing off-shell photon or $Z$ boson \cite{DrellYan1,DrellYan2} which produce a charged Higgs boson pair. As mentioned before the two modes $H^+ \to t\bar{b}$ and $H^+ \to W^+H$ produce the same final states if $H \to b\bar{b}$. The branching ratio of charged Higgs boson decay in these channels are shown in Figs. \ref{CH2tb} and \ref{CH2WH}. Assuming the minimum mass for the neutral Higgs boson as 150 GeV, charged Higgs boson masses above 250 GeV are considered so that $H^+ \to W^+H$ is kinematically allowed. 

The LHC excluded regions at 95$\%$ CL and HL-LHC exclusion expectation are shown in blue and green respectively in Figs. \ref{CH2tb} and \ref{CH2WH}. 

The light charged Higgs exclusion extends to the top mass threshold by $H^+ \to \tau\nu$ search at HL-LHC in types 2 and 4 where the leptonic decay is relevant. The heavy charged Higgs region at high $\tan\beta$ values will be covered by $H^+ \to \tau\nu$ search in type 2 and $H^+ \to t\bar{b}$ in type 3 (in this type the leptonic decay is suppressed at high $\tan\beta$). The very low $\tan\beta$ regions show normal extensions of the three analyses a, b, c when HL-LHC data is available. Therefore the region $\tan\beta\gtrsim 3$ is out of HL-LHC reach in type 1, while in type 3, the allowed region will be $3\lesssim\tan\beta\lesssim13$. 

As seen from Figs. \ref{CH2tb} and \ref{CH2WH}, in 2HDM type 1, at $\tan\beta=10$, $H^+ \to t\bar{b}$ contributes 15$\%$ to 40$\%$ of the charged Higgs decay and the rest of 60$\%$ to 85$\%$ belongs to $H^+ \to W^+H$. In type 3, the situation is in favor of $H^+ \to t\bar{b}$ absorbing 45$\%$ to 73$\%$ of the total decay fraction while the rest of 27$\%$ to 55$\%$ is taken by $H^+ \to W^+H$. Other types of the model are not considered due to neutral Higgs exclusions as shown in Fig. \ref{HExclusion}.

The charged Higgs loop contribution in the $h_{125}\to \gamma\gamma$ decay rate is also evaluated by passing decay rates from \texttt{2HDMC} to \texttt{HiggsSignals} for the model $\chi^2$ evaluation. The analysis shows preference of heavy charged Higgs bosons as discussed in \cite{CH2WH}. The $h_{125} \to \gamma\gamma$ involves $\gamma H^+H^-$ coupling which is the electric charge and $h_{125}H^{+}H^{-}$ coupling which is $-\frac{1}{v}[m^2_{h_{125}}+2(m_{H^{\pm}}^2-m^2)]$ at the alignment limit with $m^2=m_{12}^2/(\sin\beta\cos\beta)$ which is set to $m_{H}^2$ as suggested in \cite{m12} for alignment scenario. The two couplings are thus independent of the model type and $\alpha,\beta$ parameters. Taking minimum $\chi^2$ corresponding to $m_{H^+}=1$ TeV as the reference, $\Delta\chi^2$ is obtained in the charged Higgs mass and $\tan\beta$ parameter space and 68$\%$ and 95$\%$ CL bounds are drawn as vertical blue lines on the top left panel in Fig. \ref{CH2tb}. The observation is that the region of study $m_{H^+}\gtrsim250$ GeV is within the LHC bound from $h_{125}\to \gamma\gamma$ rate at 95$\%$ CL. 

The HL-LHC expectation is estimated using projected uncertainties reported by CMS and ATLAS in Tab. 35 of \cite{hllhc1} for $h_{125}$ decay channels. The uncertainties from the second scenario, S2, also called YR18, are implemented in \texttt{HiggsSignals} for the model $\chi^2$ evaluation. The lower limits for the charged Higgs mass are obtained as 700 GeV ($95\%$ CL) and 850 GeV ($68\%$ CL) as expected by HL-LHC.

\begin{figure}[hbt!]
	\centering
	\includegraphics[width=\linewidth,height=0.85\linewidth]{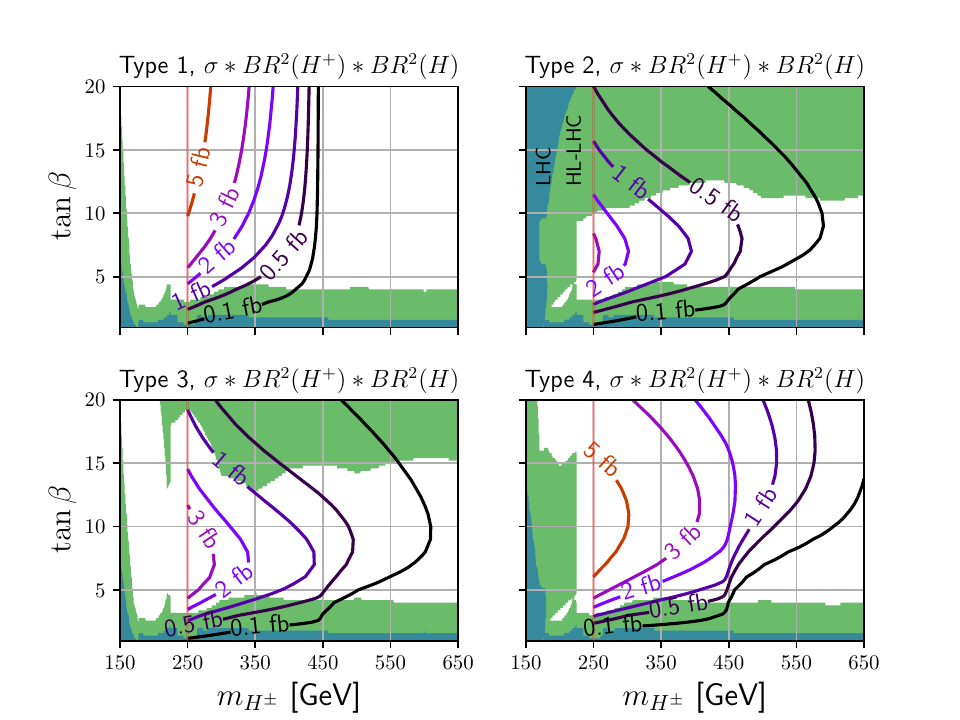}
	\caption{The cross section times branching ratio of $e^+e^- \to H^+H^- \to W^+HW^-H \to W^+W^-f\bar{f}f\bar{f}$ ($f=b$ in types 1 to 3, and $\tau$ in type 4) as a function of the charged Higgs mass and $\tan\beta$ in different types of 2HDM.}
\label{sxbr}
\end{figure}

Considering the charged Higgs boson pair production at $e^+e^-$ collisions, the signal cross section is multiplied by square of BR$(H^+ \to W^+H)*$BR$(H \to f\bar{f})$ as shown in Fig. \ref{sxbr}. Here, $f$ is $b$ in types 1 to 3, and $\tau$ in type 4. In type 1, $\sigma*$BR values decrease to below 0.1 $fb$ at masses above 450 GeV due to smallness of BR$(H\to b\bar{b})$. The observed patterns in the four types reflect the fact that there is a mass splitting between the charged and neutral Higgs bosons allowing $H^+ \to W^+H$ decay.   

\subsection{Search strategy}
\subsubsection{Cut based analysis}
\begin{figure}[hbt!]
	\centering
	\begin{subfigure}{.23\textwidth}
		\includegraphics[width=\linewidth,height=.75\linewidth]{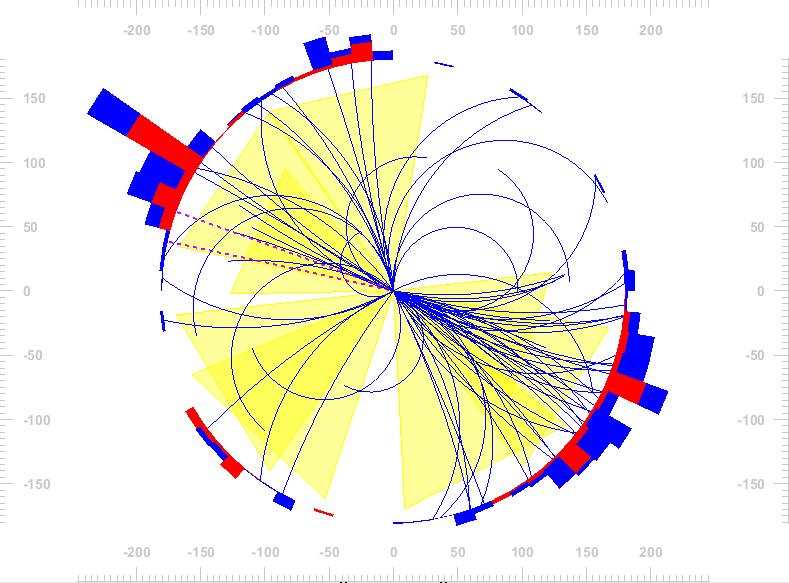}
	\end{subfigure}%
	\begin{subfigure}{.23\textwidth}
		\includegraphics[width=\linewidth,height=.75\linewidth]{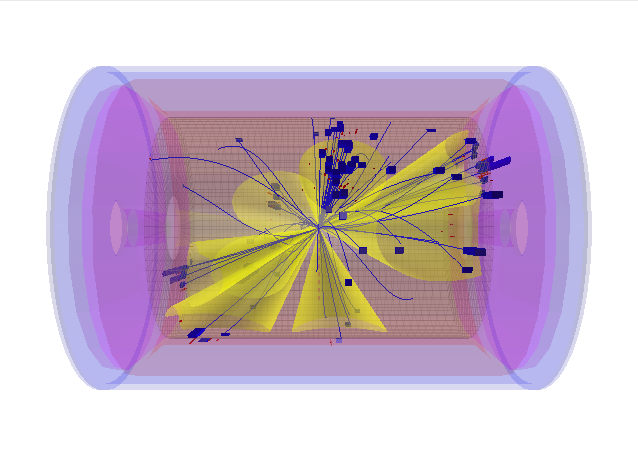}
	\end{subfigure}
		\caption{A signal event containing 8 reconstructed jets in two views of the cylindrical coordinates.}
\label{event3d}
\end{figure}

\begin{figure}[hbt!]
	\centering
		\includegraphics[width=0.6\linewidth,height=.5\linewidth]{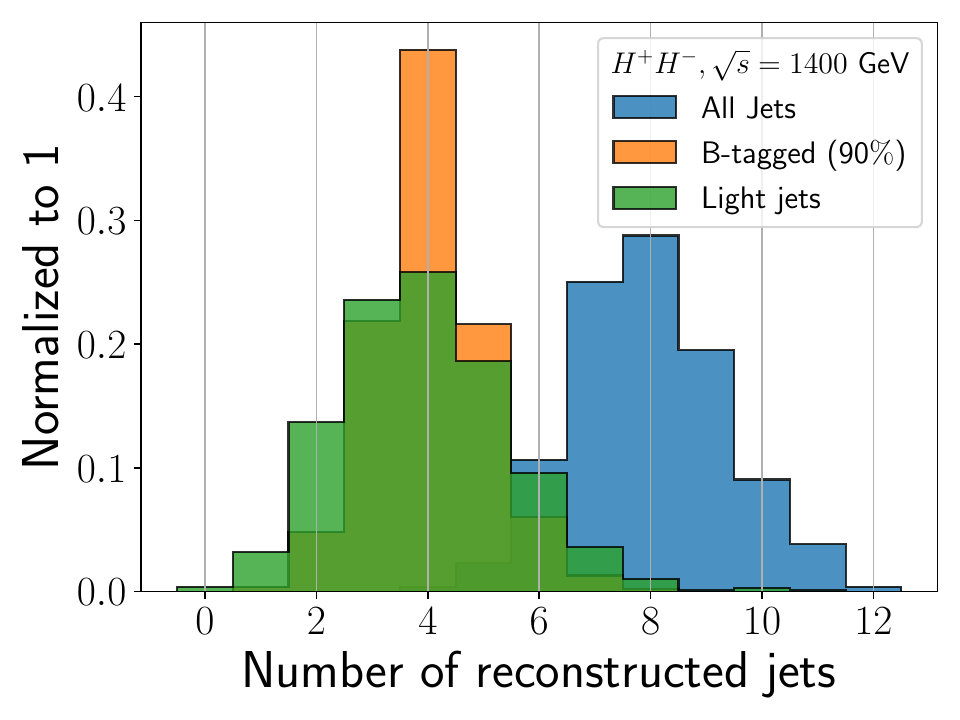}
		\caption{Number of $b$-jets, light jets and total number of jets in signal events.}
\label{SJetMul}
\end{figure}

\begin{figure}[hbt!]
	\centering
	\begin{subfigure}{.23\textwidth}
		\includegraphics[width=\linewidth,height=.85\linewidth]{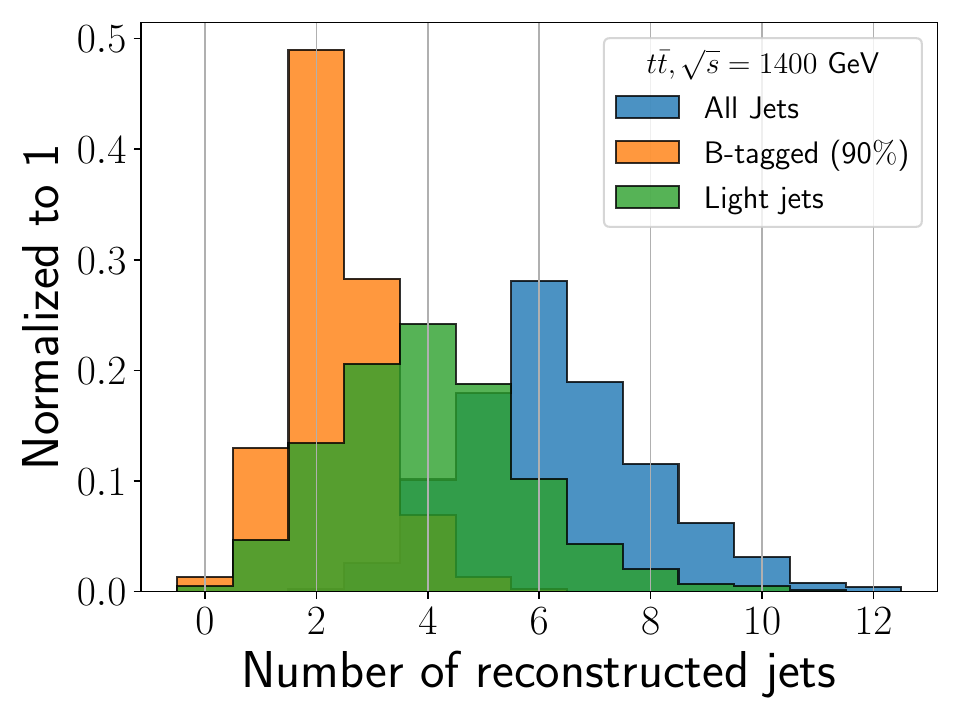}
		\caption{}
		\label{ttJetMul}
	\end{subfigure}%
	\begin{subfigure}{.23\textwidth}
		\includegraphics[width=\linewidth,height=.85\linewidth]{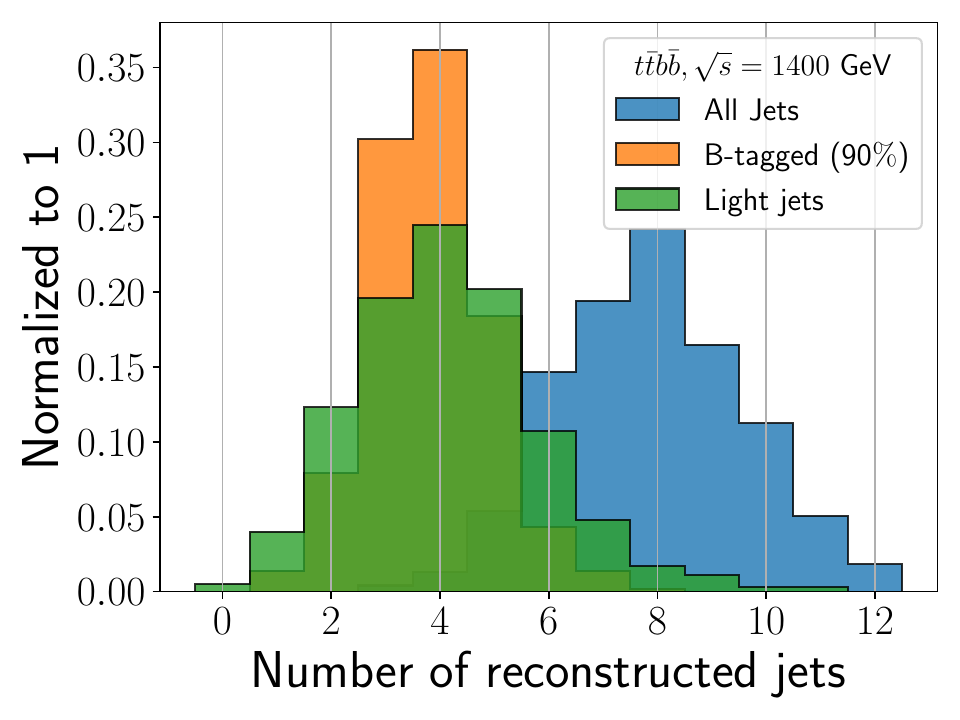}
		\caption{}
		\label{ttbbJetMul}
	\end{subfigure}
	\caption{Number of $b$-jets, light jets and total number of jets in $t\bar{t}$ (a) and $t\bar{t}b\bar{b}$ events (b).}
\end{figure}

The signal in fully hadronic final state contains a total number of eight jets. Figure \ref{event3d} shows a typical signal event in two views of a general detector concept. For event analysis, we require that there are at least eight reconstructed jets in the event using the Valencia jet reconstruction algorithm \cite{VLC1,VLC2}. The kinematic acceptance for jets is 
\begin{equation}
	p_T>5 ~\GeV ~\tn{and} ~|\eta|<5
\end{equation} 
with the usual transverse momentum and pseudorapidity definitions. A jet smearing is applied to mimic the $\gamma\gamma \to \textnormal{hadrons}$ overlay as follows:
\begin{equation}
	\frac{\delta p}{p}= \Biggl \{ 
	\begin{aligned}
	&0.01,~ |\eta| < 0.76&  \\
	&0.05,~ |\eta|\geq 0.76,& \\
	\end{aligned}
\label{smearing}
\end{equation}
i.e., $1\%$ and $5\%$ smearing with Gaussian profile is applied on the jet four-momentum in the corresponding $\eta$ regions. This is based on CLIC collaboration early proposal \cite{overlay}. The distributions of jet multiplicity are shown in Figs. \ref{SJetMul} (for the signal events with $(m_{H^{\pm}},m_{H})$= (350,150) GeV as an example) and \ref{ttJetMul}, \ref{ttbbJetMul} for $t\bar{t}$, $t\bar{t}b\bar{b}$ background events respectively. 

As seen, in average, there are equal number of light and $b$-jets in signal events leading to total number of eight jets. In the $t\bar{t}$ background events, in average, there are two $b$-jets from the top quark decay and four light jets from $W^{\pm}$ decays leading to total number of six jets. The $t\bar{t}b\bar{b}$ background produces similar jet multiplicities as in signal events due to the existence of 4 $b$-jets and total number of 8 jets in the final state.  

\begin{figure}[hbt!]
	\centering
	\begin{subfigure}{.23\textwidth}
		\includegraphics[width=\linewidth,height=.85\linewidth]{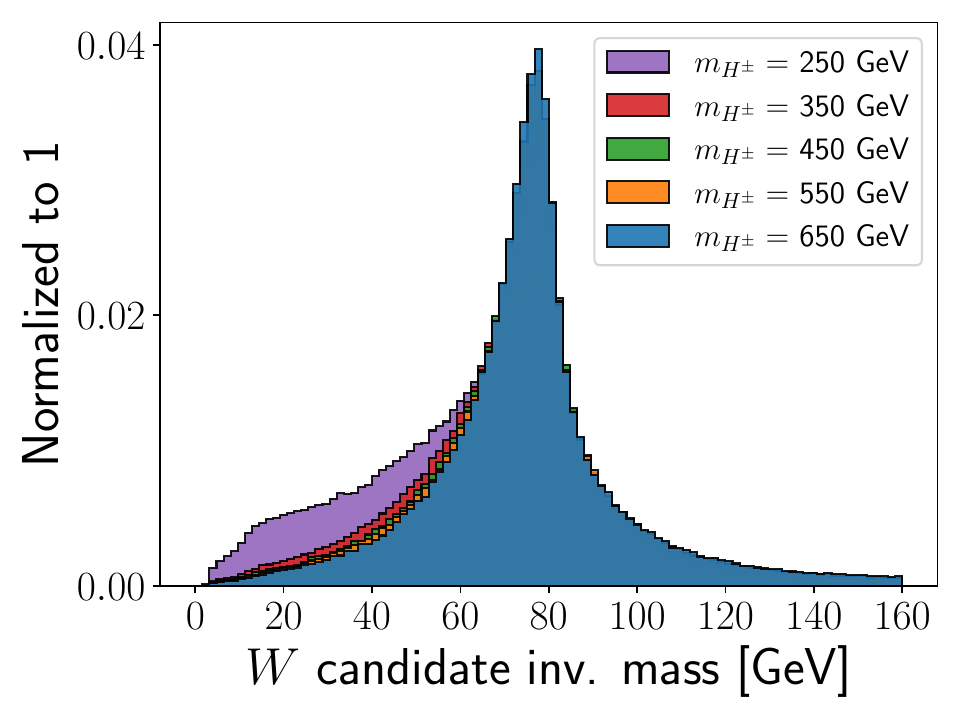}
		\caption{}
	\end{subfigure}%
	\begin{subfigure}{.23\textwidth}
		\includegraphics[width=\linewidth,height=.85\linewidth]{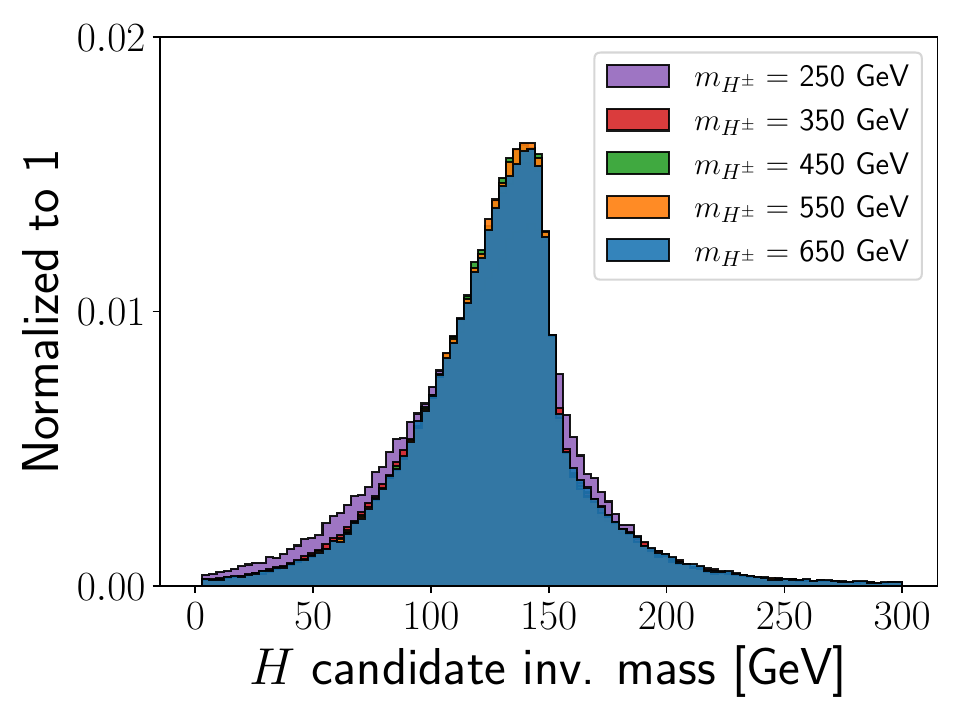}
		\caption{}
	\end{subfigure}
\caption{Invariant mass distributions of the reconstructed $W$ bosons (a) and $H$ bosons with $m_H=200~\GeV$ (b) in signal events producing different charged Higgs boson masses.}
\label{WH}
\end{figure}

\begin{figure}[hbt!]
	\centering
	\includegraphics[width=0.65\linewidth,height=.55\linewidth]{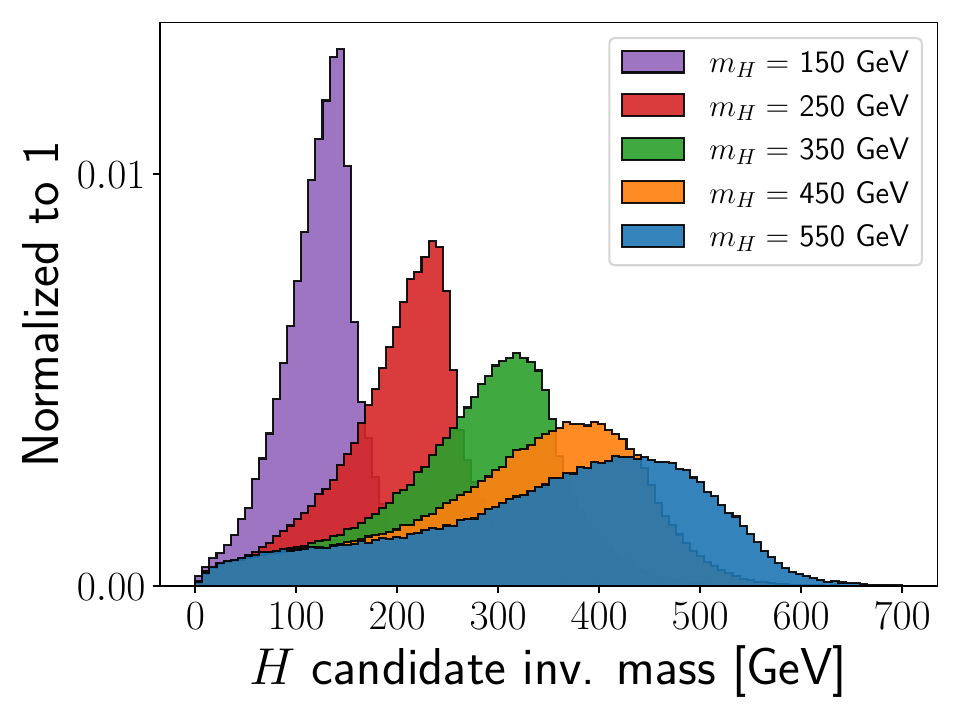}
	\caption{The neutral Higgs boson invariant mass distributions for $150~\GeV<m_H<550~\GeV$.}
\label{Hs}
\end{figure}

In the next step, the $W^{\pm}$ reconstruction is performed by searching for the correct pairs of light jets which minimize the following $\chi$:
\begin{equation}
\chi=|m_{ij}-m_W|+|m_{kl}-m_W|,
\label{chiW}
\end{equation}
where $i, j, k, l$ are light jets indices, $m_W$ is the $W^{\pm}$ boson mass set to 80 GeV and $m_{ij}$, $m_{kl}$ are invariant masses of the two pairs of light jets. 
When the best pairs of light jets are obtained through the above requirement, their invariant masses (labeled as $m_{W1}$ and $m_{W2}$) are required to be around the nominal value of the $W$ boson mass within 20 GeV window, i.e., 
\begin{equation}
	|m_{W1}-m_W|<20~ \GeV ~\tn{and}~|m_{W2}-m_W|<20~\GeV.
\label{Wmasswindow}
\end{equation}
The procedure for heavy neutral Higgs boson reconstruction is very similar to that used for $W$ boson except that the search proceeds among $b$-jets with indices $i,j,k,l$ minimizing $\chi$ defined as:
\begin{equation}
	\chi=|m_{ij}-m_H|+|m_{kl}-m_H|,
\label{chiH}
\end{equation}
followed by the mass window:
\begin{equation}
	|m_{H1}-m_H|<m_H/4~ \tn{and}~|m_{H2}-m_H|<m_H/4.
\label{Hmasswindow}
\end{equation}
The $\chi$ minimization and subsequent mass window for the heavy neutral Higgs boson relies on the knowledge of its mass, $m_H$. Therefore it is assumed that this particle has already been observed at the time of search for the charged Higgs boson through this analysis. 
Contrary to the case of $W$ reconstruction, a dynamic mass window has been used for $H$ reconstruction due to different mass hypotheses. The mass hypothesis here will be replaced by experimental input in case of successful observation of heavy neutral Higgs boson. The invariant mass distributions of reconstructed $W/H$ bosons are shown in Fig. \ref{WH} before the mass windows are applied. The neutral Higgs boson invariant mass distribution becomes wider for higher masses as shown in Fig. \ref{Hs}.   	

At this step, there are two pairs of light jets passed Eqs. \ref{chiW}, \ref{Wmasswindow} with their invariant masses denoted as $m^W_{IJ}$ and $m^W_{KL}$. The corresponding pairs of $b$-jets have passed Eqs. \ref{chiH}, \ref{Hmasswindow} with $m^H_{IJ}$ and $m^H_{KL}$ as their invariant masses. The light and $b$-jet indices are obviously different. However, we use the same letters to avoid complexity in writing. 

Before proceeding to the charged Higgs boson reconstruction, the jet four-momenta are corrected to give the correct invariant mass equal to the nominal value of the $W^{\pm}$ or $H$ bosons. This correction can be written in the following form:
\begin{equation}
	p^{\mu}_{I,J}\to p^{\mu}_{I,J}*\frac{m_{W/H}}{m^{W/H}_{IJ}},
\end{equation}
where $\mu=0 ~\tn{to}~3$ are the four-momentum indices, $I,J$ are the jet pair indices and the correction is performed for each value of $I$,$J$ found in the previous step by $\chi$ minimization. The same correction is applied on jets with indices $K$,$L$.

\begin{figure}[hbt!]
	\centering
	\includegraphics[width=0.65\linewidth,height=.55\linewidth]{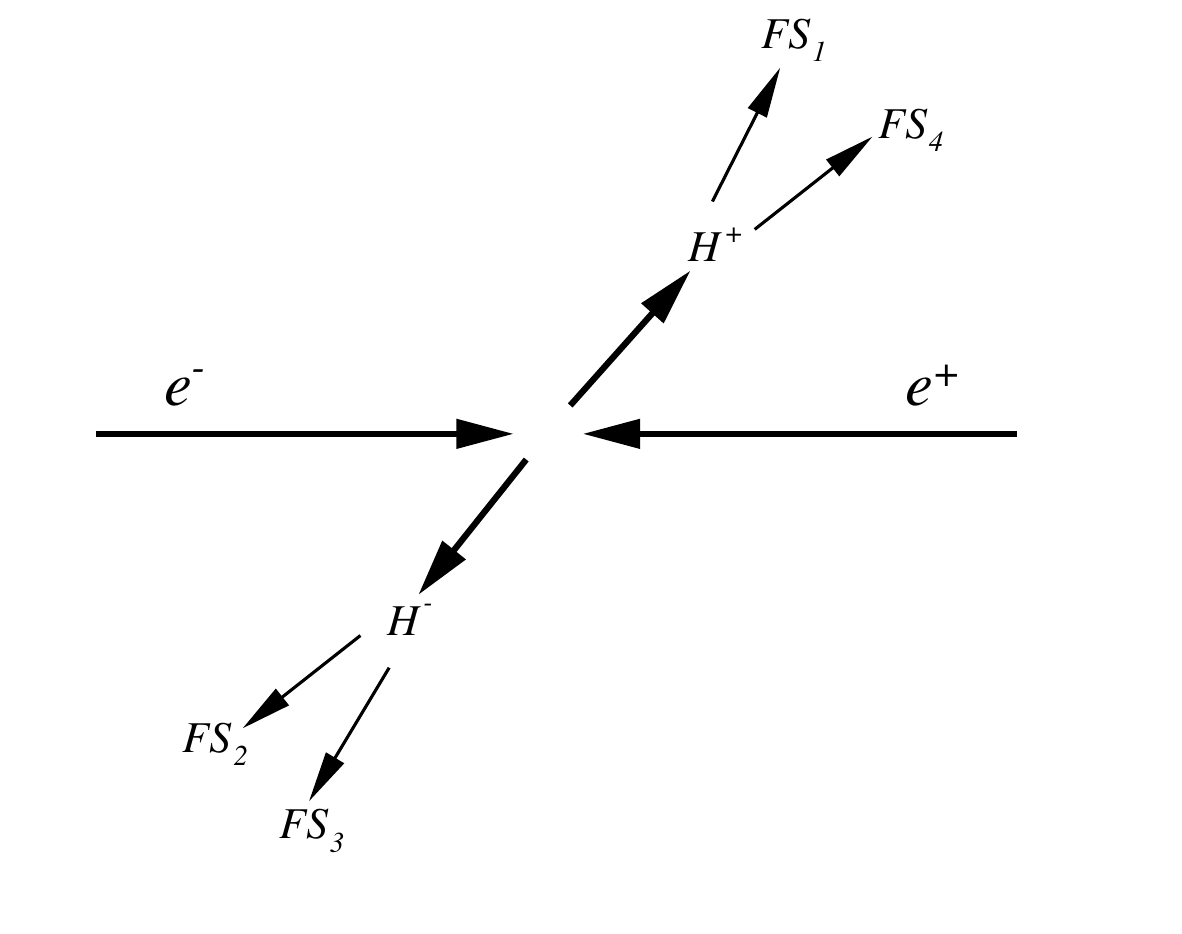}
	\caption{A signal event containing two charged Higgs bosons with four final state objects FS$_{i}$ denoting $W/H$ bosons.}
	\label{event}
\end{figure}

The corrected light and $b$-jets are then used to construct the $W/H$ four-momenta leading to $W_1,~W_2,~H_1,~H_2$ with their invariant masses equal to $m_W/m_H$ precisely. These are the four objects of the final state which can be used to reconstruct the charged Higgs bosons $H^{\pm}$. Let us call them $\tn{FS}_i$, with $i=1~\tn{to}~4$ after sorting them in descending energy. A view of signal event is shown in Fig. \ref{event}.

The pairing of the final state objects is performed by obtaining invariant masses $m_{(\tn{FS}_1,\tn{FS}_4)}$ and $m_{(\tn{FS}_2,\tn{FS}_3)}$. The idea is based on energy/momentum conservation which implies $\vec{p}_{H^{+}}=-\vec{p}_{H^{-}}$ and ${E}_{H^{+}}={E}_{H^{-}}=\sqrt{s}/2$ leading to $E_{\tn{FS}_{1}}+E_{\tn{FS}_{4}}=E_{\tn{FS}_{2}}+E_{\tn{FS}_{3}}=\sqrt{s}/2$. Therefore the final state objects with the highest and lowest energies belong to the same mother particle (charged Higgs). The indices in Fig. \ref{event} are thus selected so as to agree with the expectation.

In order to keep statistics of the signal events, we keep both combinations and fill the final invariant mass histograms with both $m_{(\tn{FS}_1,\tn{FS}_4)}$ and $m_{(\tn{FS}_2,\tn{FS}_3)}$ for double statistics.
\subsubsection{Alternative approach}
There are other methods for the signal event reconstruction. Since there are two corrected $W$ and two $H$ bosons, denoted by $W_1,~W_2,~H_1,~H_2$, the only possibilities for combining them and reconstructing the charged Higgs are $W_1+H_1\to H^+_1$, $W_2+H_2\to H^+_2$ or $W_1+H_2\to H^+_1$, $W_2+H_1\to H^+_2$. One can choose the pair of combinations which give closest charged Higgs boson masses and yield min$\Delta(m_{H^+_1},m_{H^+_2})$. This strategy was also studied leading to similar results to what already presented in the paper.  
\subsubsection{Likelihood analysis}
\begin{figure}[hbt!]
	\centering
	\includegraphics[width=0.65\linewidth,height=.55\linewidth]{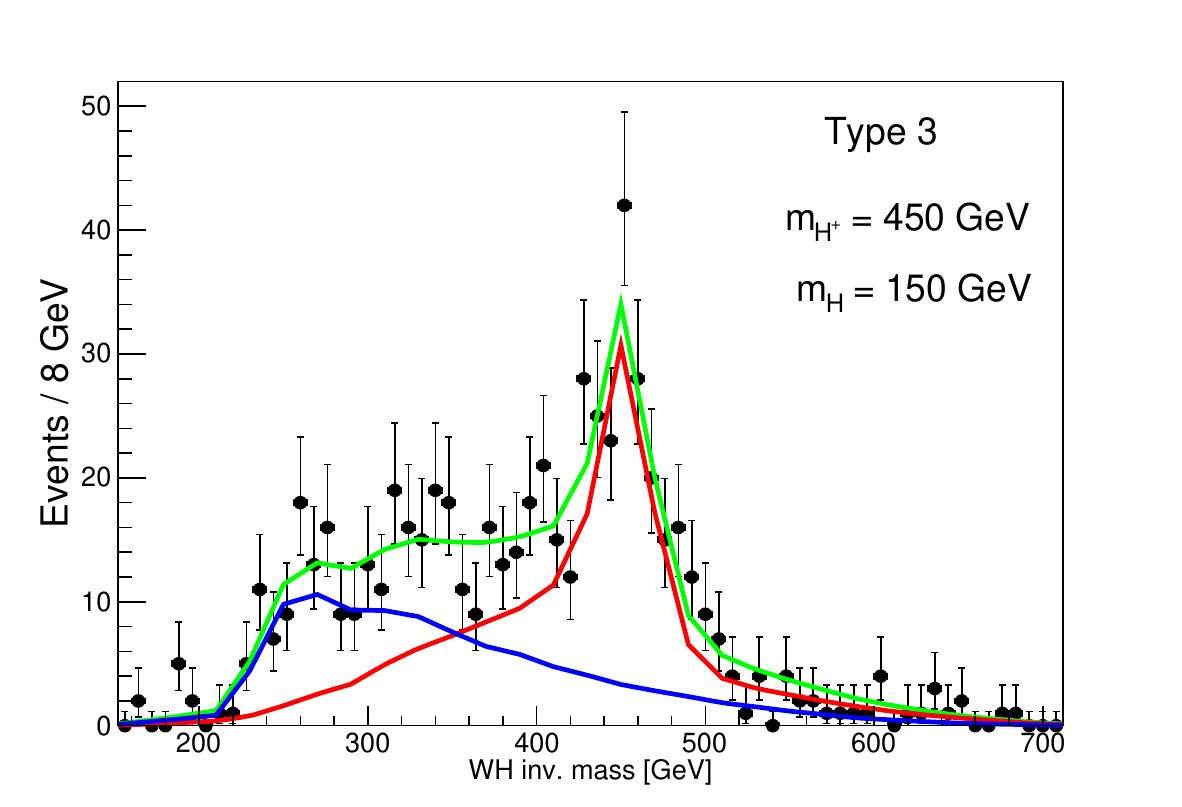}
	\caption{The pseudo-data (markers with error bars) generated based on pdfs of the signal (red line) and total background (blue line). The signal plus background pdf is shown in green.}
	\label{pseudo}
\end{figure}

The signal search strategy described here relies on cut-based counting experiment which applies dynamic mass windows on the final invariant mass distributions. An alternate approach is based on binned histogram extended likelihood analysis. In order to do so, the signal and background (invariant mass) histograms are used to get probability density functions (pdfs) of the distributions. Based on these pdfs, pseudo-data are generated and a profile likelihood is calculated for hypothesis tests. The signal significance is then obtained by comparing the null hypothesis (background only, no signal events) with the alternate hypothesis of signal plus background. Example of a signal ($(m_{H^+},m_H)=(450,150)~\GeV$ in type 3) and background model and generated data are shown in Fig. \ref{pseudo}.  
      
\section{Results}
\begin{figure*}
	\centering
	\includegraphics[width=\linewidth,height=0.8\linewidth]{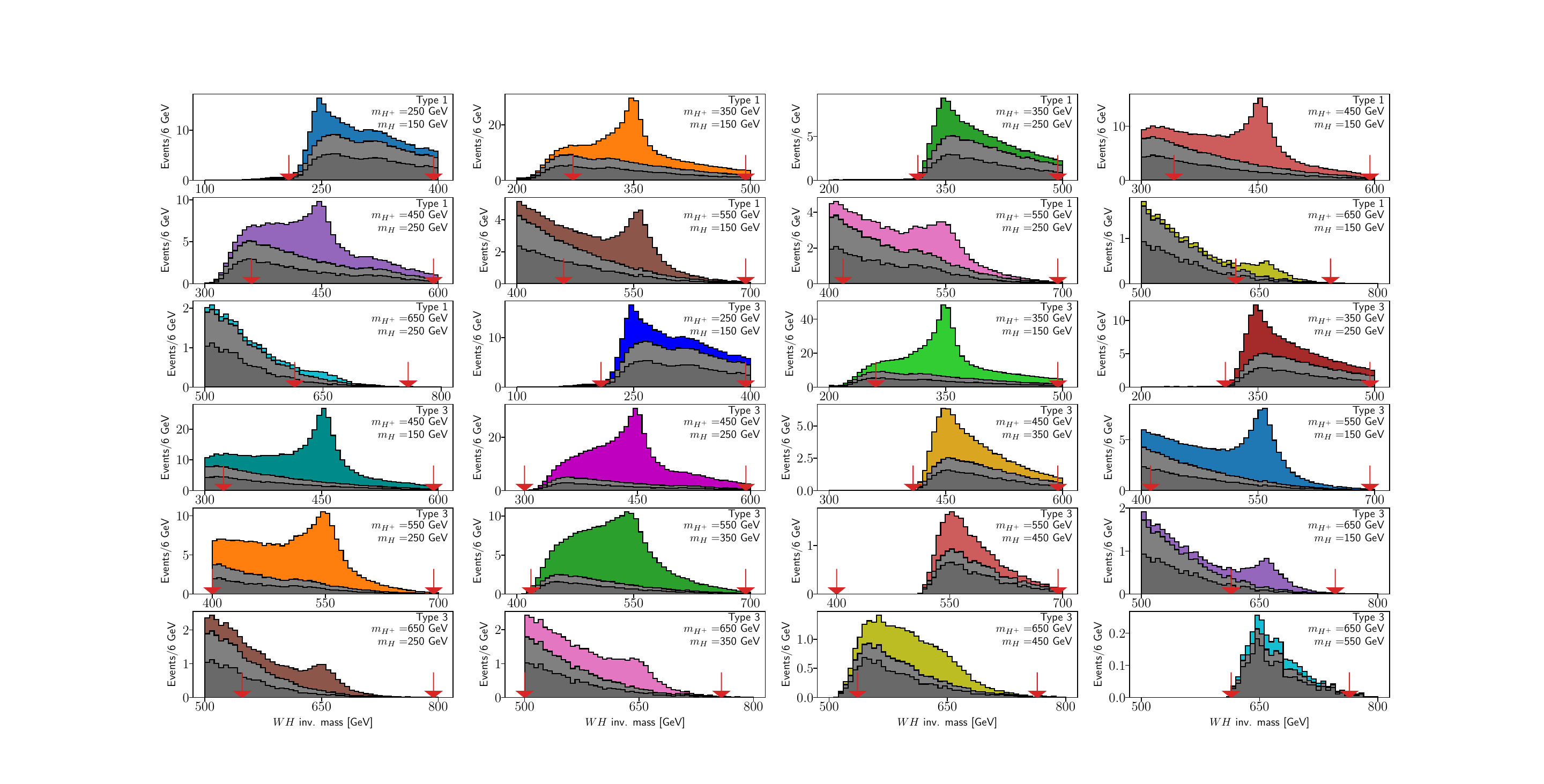}
	\caption{The charged Higgs candidate invariant mass in 2HDM types 1 and 3 with different mass assumptions at 1000 $fb^{-1}$.}
	\label{SBAll}
\end{figure*}

\begin{figure}[hbt!]
	\centering
	\includegraphics[width=0.8\linewidth,height=.7\linewidth]{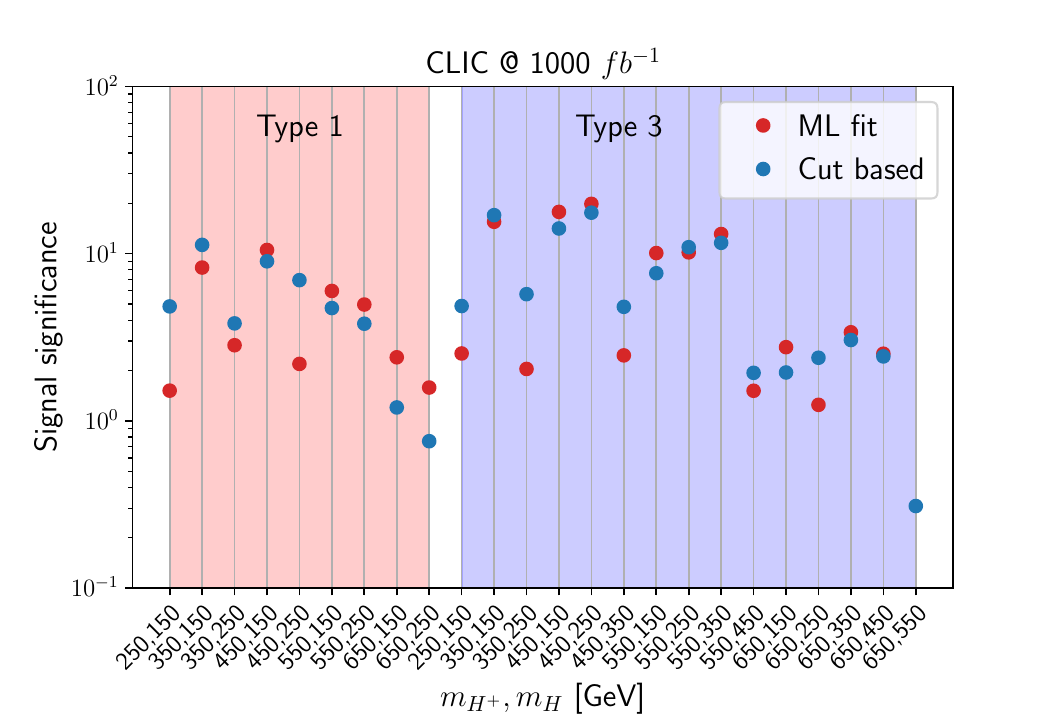}
	\caption{Comparison of the signal significances obtained from the cut based analysis and the binned maximum likelihood fit.}
	\label{signif}
\end{figure}

Figure \ref{SBAll} shows the charged Higgs invariant mass distributions in 2HDM types 1 and 3 with different charged and neutral Higgs boson mass hypotheses. The value of $\tan\beta$ is set to $\tan\beta=10$. All distributions are normalized to integrated luminosity $\mathcal{L}=1000~fb^{-1}$. The $t\bar{t}$ background (shown in dark gray) has a total cross section of 145 $fb$. The $t\bar{t}b\bar{b}$ background which contains $t\bar{t}Z$ and $t\bar{t}h_{125}$ followed by $Z/h_{125}\to b\bar{b}$ is shown in light gray and has a cross section of 3.8$fb$. Other SM backgrounds ($Z/\gamma$, $ZZ$, $W^+W^-$, $b\bar{b}b\bar{b}$) have no contribution after jet multiplicity and $b$-tagging requirements. 
The red arrows show the mass window which provides the maximum signal significance defined as $N_S/\sqrt{N_S+N_B}$ where $N_S(N_B)$ is the signal(background) number of events in the final histogram within the mass window. 
The signal significance results for both cases of cut based and likelihood analyses are shown in Fig. \ref{signif}.

In 2HDM type 1, as Fig. \ref{sxbr} indicates, the signal cross section times branching ratio of charged and neutral Higgs boson decays is almost independent of $\tan\beta$ for $\tan\beta \geqslant 10$. Therefore results of the type 1 at $\tan\beta=10$ are valid for higher $\tan\beta$ values. On the other hand, the neutral Higgs boson decay to $t\bar{t}$ which is activated at $m_H \geqslant 2m_t$, limits the region of parameter space to large mass splitting above 100 GeV for $m_{H^+}> 450 ~\GeV$ as $m^\tn{max}_H\simeq 350~\GeV$ for $H\to b\bar{b}$. Possibility of $H^+\to W^+H$ with $H\to t\bar{t}$ should then be explored using top tagging techniques based on jet substructure \cite{BDRS} demanding four (fat) top jets to be found in the event \cite{tt1,JohnHopkins,tt3}. Due to the top quark jet substructure in hadronic decay mode and the large jet multiplicity in signal events (16 jets), exploring the region of $m_{H^+}>450~\GeV$ and $m_H>350~\GeV$ is challenging in 2HDM type 1.  

In 2HDM type 3, a broader region is accessible with $\sigma\times \tn{BR}$ decreasing as $m_{H^+}$ increases. There is $\tan\beta$ dependence in final results in this case and the best region is near $\tan\beta=\sqrt{m_t/m_b}\simeq 6.5$ which is where $H^+ \to t\bar{b}$ is suppressed as mentioned before. The $\tan\beta$ values below or above this region lead to reduction in BR$(H^+ \to W^+H)$ resulting in less signal rate. 

Other types, are limited by neutral Higgs boson exclusions as shown in Fig. \ref{HExclusion}. In type 2, a heavy neutral Higgs boson can be used for the signal search in $H\to b\bar{b}$ mode at $\tan\beta<5$ with masses above 350 GeV, but these masses are covered partially by HL-LHC. The type 4 is suitable with $H\to \tau\tau$ again at small region of $\tan\beta<5$ which looks out of HL-LHC reach. The analysis of this type should be similar to what has already been done at LHC \cite{CMS:chnew2} keeping in mind the different nature of events and experiments.  

\begin{figure}[hbt!]
	\centering
	\includegraphics[width=0.8\linewidth,height=.8\linewidth]{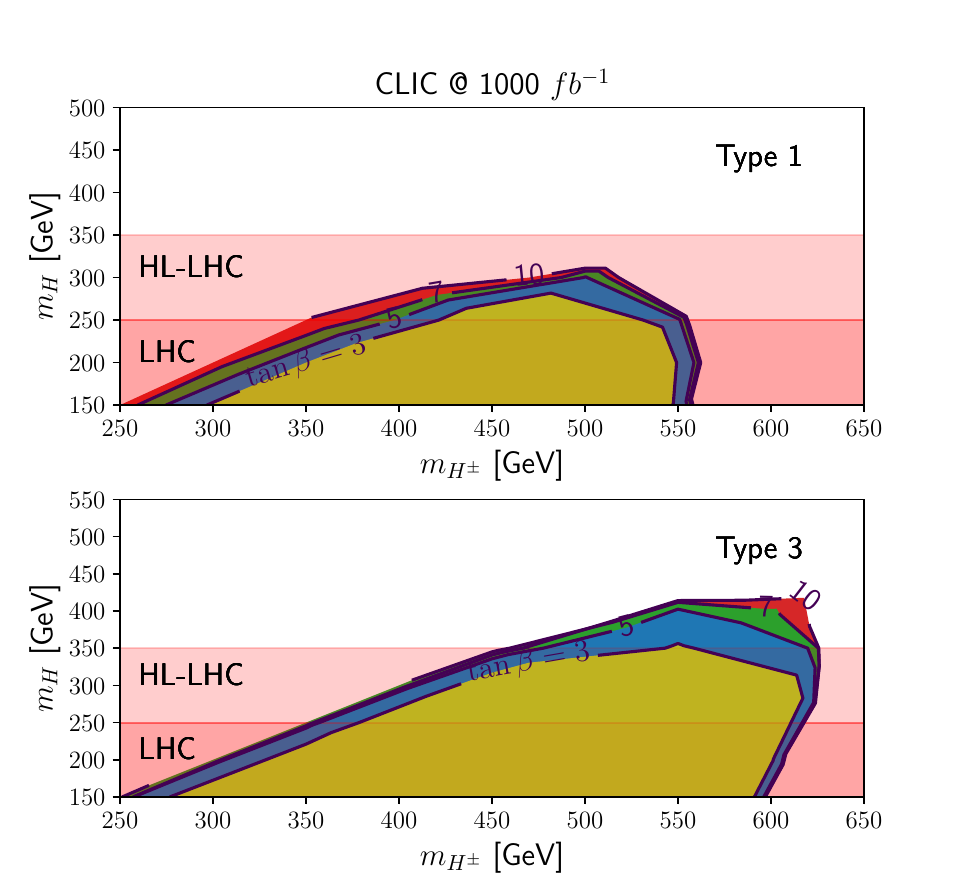}
	\caption{The $5\sigma$ contours of the 2HDM type 1 and 3 at different $\tan\beta$ values. The regions denoted as LHC and HL-LHC show the neutral Higgs excluded regions by the current LHC data and future HL-LHC expectation.}
	\label{contour}
\end{figure}

Since final results depend on the $m_{H^{+}}, m_H$ and $\tan\beta$, a scan over the parameter space of interest is performed to visualize the 5$\sigma$ contours in types 1 and 3. Results are shown in Fig. \ref{contour} as a function of the charged and neutral Higgs bosons for different $\tan\beta$ values below 10. The LHC neutral Higgs boson excluded region at 95$\%$ CL as well as HL-LHC expectation are also shown in Fig. \ref{contour}. 
       
\section{Conclusions}
A search strategy for heavy charged Higgs boson decay to $W^{\pm}H$ was described as a proposal for a high energy lepton collider. The analysis was performed assuming CLIC operation at $\sqrt{s}=1400$ GeV with fast detector simulation including parametrized $b$-tagging algorithm, jet reconstruction and momentum smearing. The beam spectrum was included and the effect of $\gamma\gamma \to \tn{hadrons}$ overlay background was added as jet momentum smearing according to CLIC collaboration studies.

The neutral Higgs boson decay through $H\to b\bar{b}$ was used together with hadronic decay of $W$ bosons to search for the signal events in fully hadronic final state. The charged Higgs masses 250 GeV to 650 GeV were studied with a minimum of 100 GeV mass difference with neutral Higgs bosons to allow $H^+ \to W^+H$ kinematically.

Results were obtained as invariant mass distributions of charged Higgs boson candidates on top of the SM background followed by a mass window optimization and likelihood analysis. 

The conclusion is that although a sizable area is achievable by lepton collider outside the current LHC excluded region, a large coverage is expected by further data from HL-LHC and a lepton collider with $\sqrt{s} = 1400$ GeV only provides a complementary probe of the parameter space but has no discovery potential in the context of the model and process discussed in the paper.
     

%

\end{document}